\documentclass[a4paper,11pt]{article}
\pdfoutput=1 

\usepackage{jinstpub} 

\usepackage{lineno}

\usepackage{caption}

\title{\boldmath STRV – A radiation hard RISC-V microprocessor for high-energy physics applications}

\author[a,1]{A. Walsemann,\note{Corresponding author.}}
\author[a]{M. Karagounis}
\author[b]{A. Stanitzki}
\author[c]{D. Tutsch}

\affiliation[a]{University
of Applied Sciences and Arts Dortmund,\\Sonnenstraße 96-100, 44139 Dortmund, Germany}
\affiliation[b]{Fraunhofer Institute for Microelectronic Circuits and Systems,\\Finkenstraße 61, 47057 Duisburg, Germany}
\affiliation[c]{University of Wuppertal,\\Gaußstraße 20, 42119 Wuppertal, Germany}

\emailAdd{alexander.walsemann@fh-dortmund.de}

\abstract{While microprocessors are used in various applications, they are precluded from the use in high-energy physics applications due to the harsh radiation present.
To overcome this limitation a microprocessor design must withstand high doses of radiation and mitigate radiation induced soft errors.
A TMR protection scheme is applied to protect a RISC-V microprocessor core against these faults.
The protection of the integrated SRAM by an independent scrubbing algorithm is discussed.
Initial irradiation results and power consumption measurements of the radiation-resistant RISC-V microprocessor implemented in 65 nm CMOS  technology are presented.
}


\proceeding{Topical Workshop on Electronics for Particle Physics 2022\\
    19-23 September, 2022\\
    Bergen, Norway}

\begin{document}
\maketitle
\flushbottom

\section{Introduction}
Microprocessors are commonly used circuits for a plethora of applications.
However, in high-energy experiments such as the ATLAS and CMS detectors installed at the Large Hadron Collider (LHC) at CERN they have not been used so far in locations that are close to the beam pipe or the interaction point. The reason for the limited use is the harsh radiation environment, which reaches a Total Ionizing Dose (TID) several orders of magnitude higher than radiation levels in space applications, where microprocessors are commonly found, and a very high Single Event Effect (SEE) rate due to a particle flux of up to 1.5 GHz / cm$^2$\cite{Conti:2312583}.
%
\textcolor{black}{The radiation found in these environments can be hazardous to microprocessors by causing critical arithmetical and functional failures.}
The use of SEE mitigation techniques is required to improve the reliability of a microprocessor core and associated system components to a level that meets the allowable residual risk of the application.

RISC-V is a natural candidate for the implementation of an open non-commercial microprocessor platform intended for fundamental research. The SEE tolerant STRV-R1 RISC-V microprocessor presented here is implemented in a 65 nm CMOS technology.
To the authors knowledge the STRV-R1 is the first microprocessor design targeting high-energy physics applications and their respective radiation levels. 
Integrating a RISC-V core in such environments could enable the faster and cheaper development of new systems compared to the traditional use of custom ASICs with fixed logic through simple \textcolor{black}{software} updates rather than extensive redesigns.
\textcolor{black}{The intended use of the STRV-R1 is to perform environmental measurements, power supply monitoring, and control applications currently performed by ASICs such as the AMAC\cite{CERN_2017} and MoPs\cite{MoPs_2020}.}
\textcolor{black}{It will serve as a baseline for future research to analyze the resource impact caused by the introduction of traditional radiation protection schemes.}

\section{RISC-V RV32IMC Core}
RISC-V is a new instruction set architecture
(ISA) that is provided under an open-source license, allowing for free use in academia and industry. Its development began in 2010 at the University of California, Berkeley\cite{RISC-V_2011}.
The specific RISC-V core implementing the ISA used in this design is based on the \textcolor{black}{Fraunhofer} AIRISC core\cite{AIRISC_2021}. The AIRISC core features the RV32I ISA variant with the full set of 32 registers and includes the M-extension for accelerated integer multiplication and division in addition to the C-extension for compressed instructions.
The pipeline has three stages for instruction fetch, decode / execute and writeback.
\textcolor{black}{The system is designed for a clock frequency of 50 MHz to provide sufficient computing power for monitoring and control applications with a limited power budget.}
Instructions and data are stored within the same 32 kB SRAM. This enables a flexible layout with an adjustable ratio between program and data memory size.
A memory bridge placed between the RISC-V core's IMEM instruction bus, it's DMEM data bus and the SRAM allows both buses to access the same SRAM, even with single-port SRAM.

\subsection{System Architecture}
The STRV-R1 is divided into three domains, which allows a finer analysis of the system power consumption in addition to more granular presence detection of single event effects.
The core domain incorporates the RISC-V core, the debug module and the memory bridge.
The SRAM domain contains the SRAM cell macros, the accompanying SRAM interface and the protection algorithm described in section \ref{sram_ref}. 
The peripherals domain provides 27 configurable GPIOs and a single UART interface.
The system also contains three additional memory mapped 32-bit counters located in the peripheral domain, which record the number of SEUs detected within each domain.
All modifications introduced in the RISC-V core adhere to the RISC-V specification and no system specific modifications of the executed software instructions are required.

\section{Radiation Hardening}
The STRV-R1 is designed to cope with high fluences and high ionizing radiation doses.
The STRV-R1 uses triple modular redundancy (TMR) to mitigate soft errors cause by single event effects.
TID effects are targeted through the use of a 65 nm fabrication technology in combination with thin gate oxide transistors which achieves an inherently high TID hardness\cite{7348757}.
The exclusive use of thin gate oxid transistor limits the system voltage including the I/O pads to a typical value of 1.2V.

\subsection{\textcolor{black}{Triple} Modular Redundancy}
Triple modular redundancy is an established fault tolerance technique for avoiding errors caused by radiation effects.
STRV-R1 uses full TMR, meaning a triplication of all combinational logic, all sequential logic elements, and the use of three majority voters, three separate clock trees and reset signals for the sequential logic.
The fine-grained TMR implemented in STRV-R1 contains a voter after every sequential element ensuring that the following combinational logic is not exposed to erroneous signals.
The prolonged presence of SEUs in a sequential element results in an increased probability of  a non-correctable error, caused by two SEUs within the same instance group.
\mbox{Figure \ref{fig_tmr}} shows the integration of a feedback path that updates sequential elements when no new data is stored to counteract this issue.
The SRAM in the STRV-R1 is protected by applying TMR to the memory.
For this purpose three SRAM blocks are used which have their output majority voted by three 32 bit wide voters before the data is transferred to the RISC-V core.
During a write access the data contained in a TMR buffered register is written to the SRAM.

The majority voters have an additional output that provides a signal when a discrepancy is detected between the three instances. These signals are passed via or-gates to a counter.
The value of the counter can be accessed via a memory mapped register from the RISC-V core.

\subsection{SRAM Refresh}
\label{sram_ref}
Accumulations of SEUs in the SRAM is a major concern, since there is no upper time limit between write accesses to a particular row of the memory. This applies particularly to the data in the instruction section of the memory layout which is likely only written once directly after startup. 
To oppose this issue STRV-R1 relies on a self-refresh algorithm that periodically refreshes the SRAM content with corrected data. This scrubbing algorithm is fully independent from the RISC-V core.
It is designed to contain only a minimal amount of logic, resulting in a small SEE cross section. The refresh algorithm nevertheless still utilizes TMR itself to guard against SEEs. The system uses dual-port SRAMs to provide the RISC-V core and the self-refresh algorithm with independent access to SRAM content.
\begin{figure}[htbp]
	\centering
	\begin{minipage}{.5\textwidth}
		\centering
		\includegraphics[width=\columnwidth]{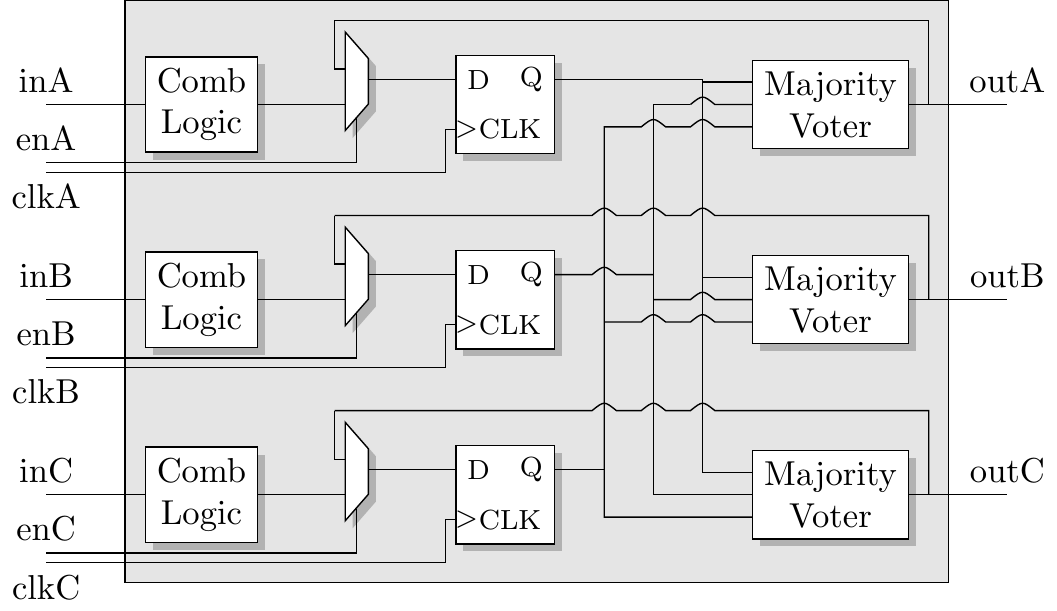}
		\caption{TMR structure with feedback path}
		\label{fig_tmr}
	\end{minipage}
	\begin{minipage}{.49\textwidth}
		\centering
		\includegraphics[width=0.47\columnwidth]{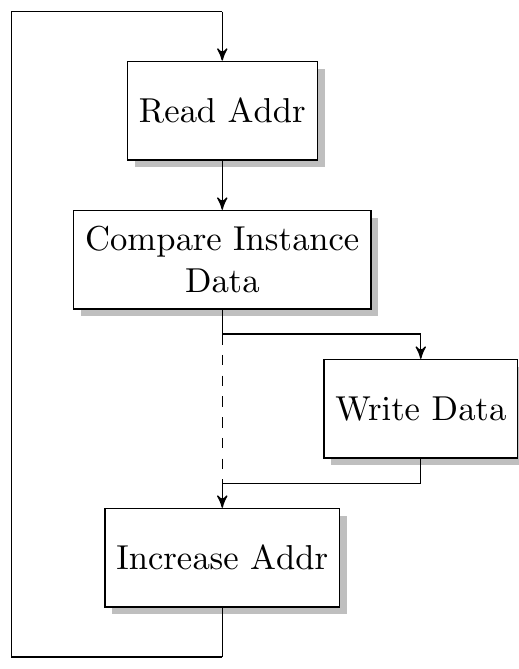}
	\caption{SRAM refresh algorithm}
	\label{fig_sram_ref}
	\end{minipage}%
\end{figure}
Fig. \ref{fig_sram_ref} illustrates the proposed self-refresh scheme. The data in each row of the SRAM is read and checked periodically. If an upset in any bit of the data provided by the three SRAM instances is detected, the three 32 bit words are majority voted. The majority voted data is then written back to SRAM during the next clock cycle. If the data is matching or the RISC-V core is writing to the same row the write access is skipped and the next row of the SRAM is read. 
The scrubbing enforces a hard limit on the time frame in which each row contains uncorrected soft-errors.
\section{Measurement Results}
All measurements in section \ref{sec:power} are done on a non irradiated chip.
The required clocks and supply voltages for the operation of the chip are provided by external sources.
The chip is programmed through the JTAG interface.
Running the Dhrystone 2.1 benchmark the system achieves 0.628 DMIPS / MHz. Using the O3 GCC compiler option yields a value of 0.665 DMIPS / MHz.

\subsection{Power Consumption}
\label{sec:power}
The power consumption is measured using an external digital multimeter for each of the three domains. An aperture time of 200 ms is selected for each conversion which captures the average current consumption over several benchmark iterations.
Only the actual execution of the benchmark is considered during the power consumption measurement. The startup of the chip and programming via JTAG are excluded.
\begin{figure}[ht]
	\begin{minipage}{.5\linewidth}
		\includegraphics[width=0.84\textwidth]{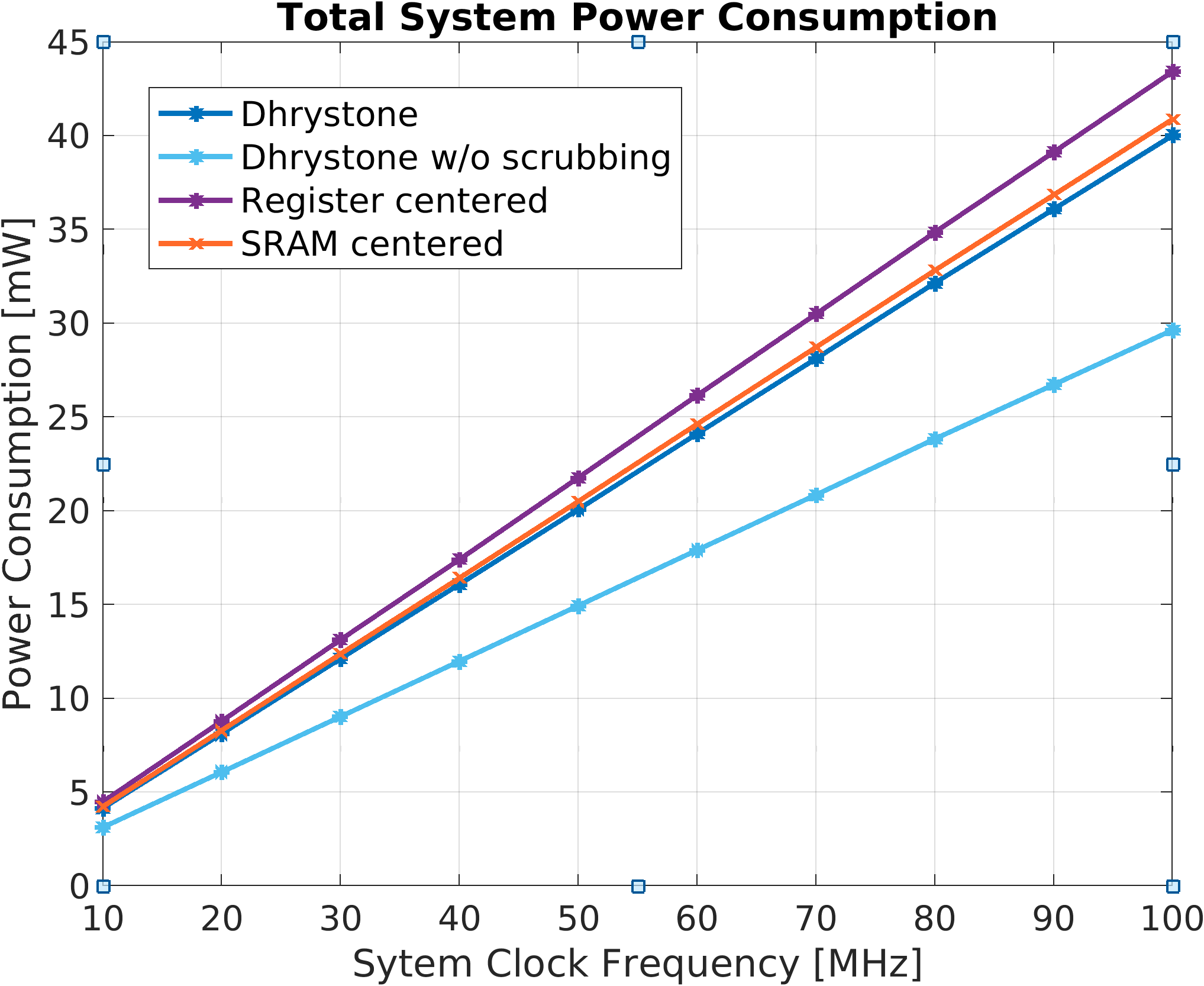}
		\caption{Total system power consumption}
		\label{fig_power_clk}
	\end{minipage}
	\hfill
	\begin{minipage}{.49\linewidth}
			\captionof{table}{Power consumption of each domain}
			\label{tab_meas}
			\resizebox{\textwidth}{!}{%
				\begin{tabular}{|l|c|c|c|c|}
					\hline
					&\multicolumn{4}{c|}{\textbf{Power Consumption [mW]}} \\
					\multicolumn{1}{|c|}{\textbf{Scenario}} & \textbf{\textit{Core}}& \textbf{\textit{SRAM}} & \textbf{\textit{Peripherals}} & \textbf{\textit{Total}}\\
					\hline
					Dhrystone Benchmark$^{\mathrm{a}}$	& 7.33 & 10.26 & 2.49 & 20.08\\
					Register centered$^{\mathrm{b}}$ 	& 8.55 & 10.68 & 2.55 & 21.77\\
					SRAM centered$^{\mathrm{c}}$		& 7.61 & 10.44 & 2.45 & 20.51\\
					$^{\mathrm{a}}$ w/o SRAM refresh	& 7.33 & 5.12 & 2.49 & 14.94\\
					$^{\mathrm{b}}$ w/o SRAM refresh	& 8.54 & 5.54 & 2.55 & 16.63\\
					$^{\mathrm{c}}$ w/o SRAM refresh	& 7.61 & 5.30 & 2.46 & 15.37\\
					\hline
					\multicolumn{4}{l}{Measured with 1.2 V supply and at 25 °C.}
				\end{tabular}
			}
	\end{minipage}
\end{figure}
Table \ref{tab_meas} shows the measured power consumption for several instrcution scenarios executed at 50 MHz.
The register centered scenario contains a majority of instructions that address the general-purpose registers. The SRAM centered scenario in contrast primarily executes load and store operations resulting in SRAM memory access. The higher power consumption in the core domain observed during the register centered scenario is caused by the absence of pipeline stalls, since there is no requirement to wait for an SRAM access to complete. Little variance in the SRAM power consumption between the different scenarios has been observed, since the access to the SRAM is permanently utilized by either the instruction bus or the data bus. From these measurements, it can be concluded that the SRAM macros and their SEU mitigation strategy are a dominant factor in the power consumption of the system.
No change in the power consumption of the other two domains is measured, when the SRAM refresh algorithm is enabled.

Figure \ref{fig_power_clk} shows the linear dependency between the power consumption and the frequency.
The system power consumption varies between 268 \textmu W / MHz and 300 \textmu W / MHz depending on the type of instructions executed. 
The SRAM refresh algorithm requires an additional 88 \textmu W / MHz independent of the instructions executed. The measured leakage power of the system is 110 \textmu W.

\subsection{X-ray Irradiation}
\begin{figure}[htbp]
    \begin{minipage}{.49\textwidth}
      \includegraphics[width=0.95\linewidth]{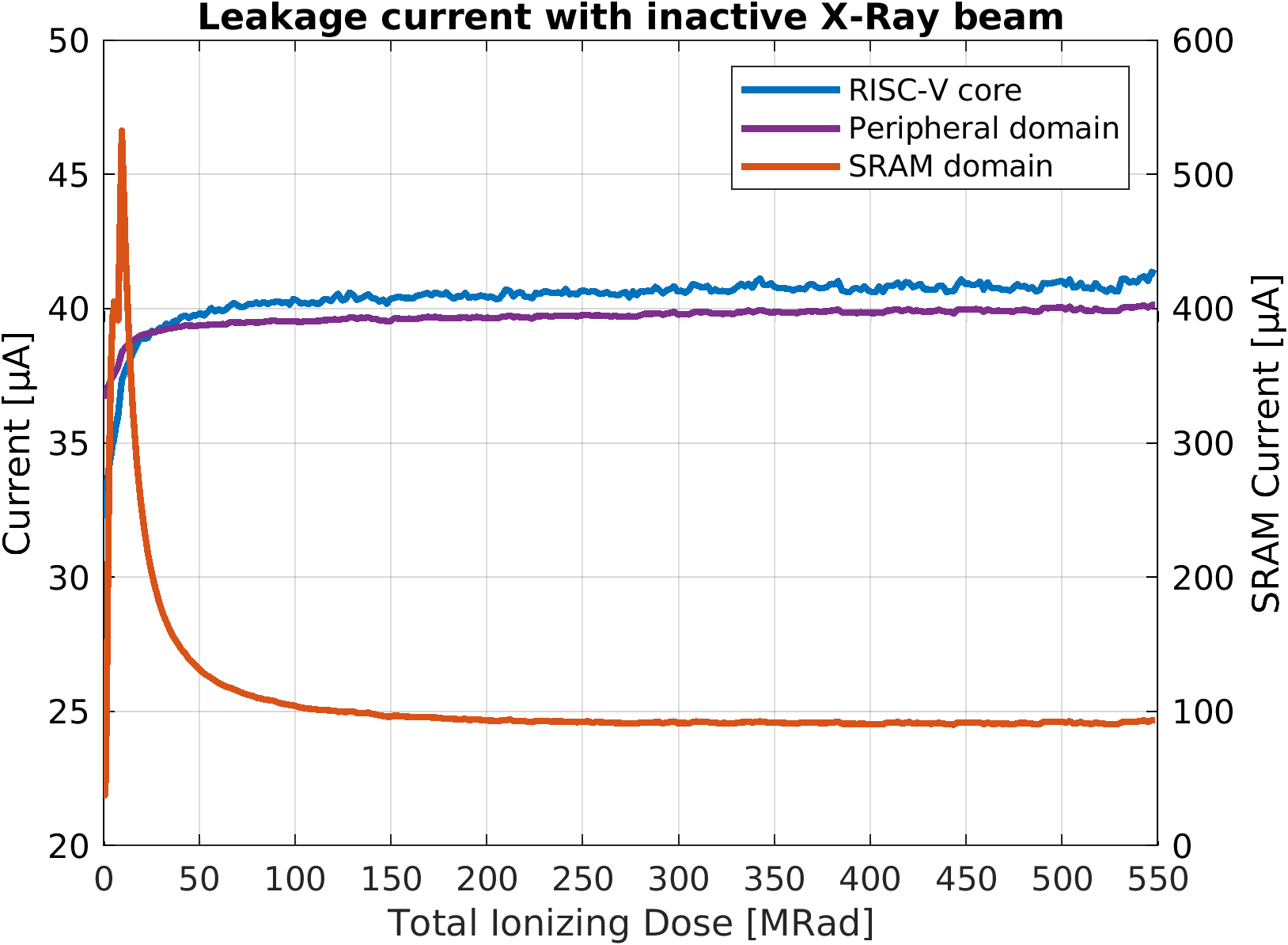}
      \caption{Leakage currents of each domain\\ as a function of TID at 1,2 V and -10 °C}
      \label{fig_leakage}
    \end{minipage}%
    \begin{minipage}{.49\textwidth}
      \includegraphics[width=0.95\linewidth]{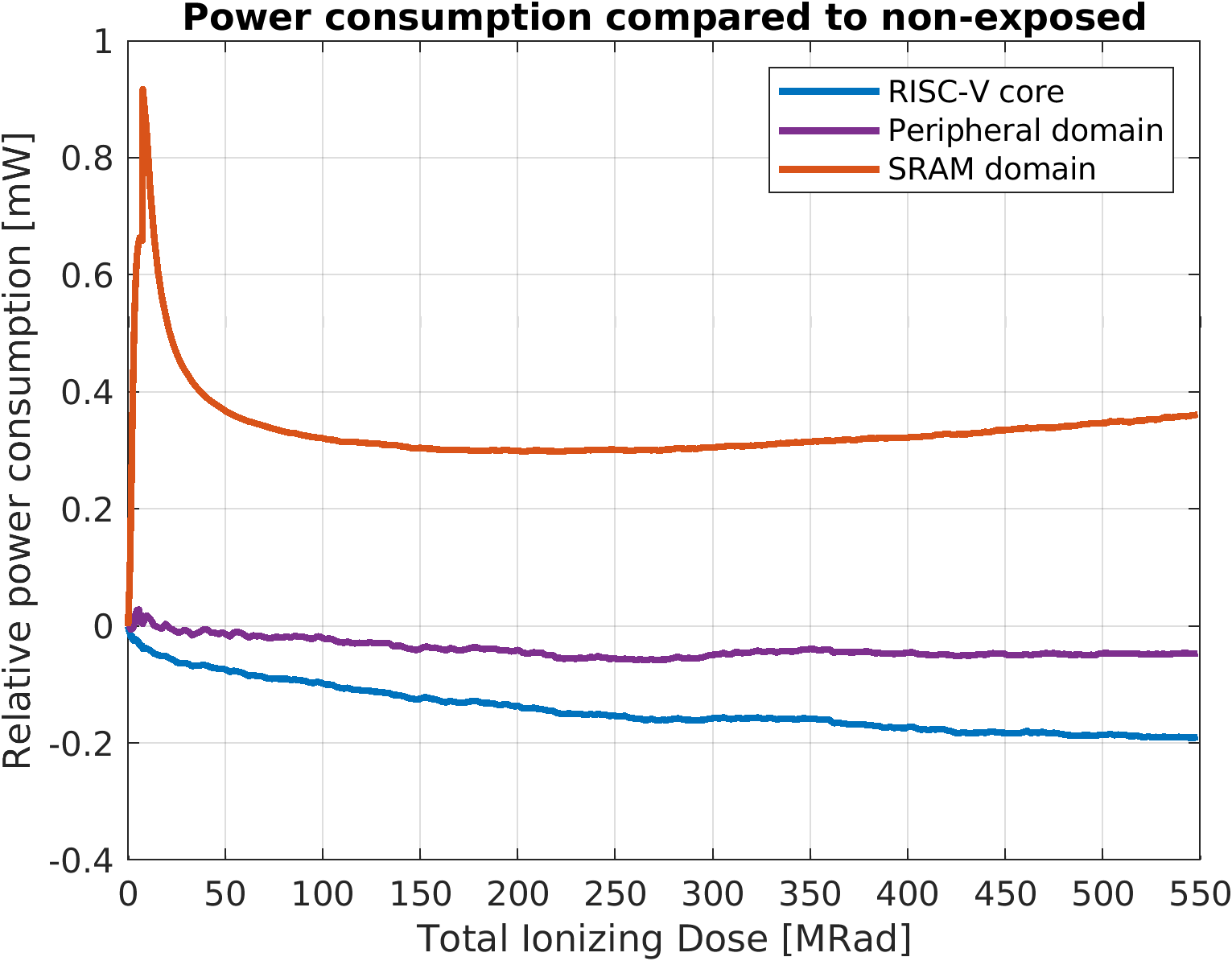}
      \caption{Relative system power consumption\\ as a function of TID  at 1,2 V and -10 °C}
      \label{fig_power}
    \end{minipage}
\end{figure}
To verify the hardness of the STRV-R1 against accumulative effects it has been irradiated to at least 500 Mrad of Total Ionizing Dose.
\textcolor{black}{The irradiation was performed with X-rays covering a spectrum up to 50 keV at a dose rate of \mbox{1.75 Mrad / h} and was carried out at two different temperatures of -10 °C and -40 °C.}
No permanent or temporary functional disruption was observed in the core or SRAM cells throughout the irradiation. System instabilities during operation indicating a critical timing deviation in one of the components were also not observed.
Figure \ref{fig_leakage} shows the leakage current for each of the domains at -10 °C. The core domain has shown an increase in leakage current of 9.2 \textmu A (28 \%).
A major increase in the leakage current of the SRAM from 37.6 \textmu A to 561.8 \textmu A (1490 \%) was measured, peaking at a dose of 8.5 MRad.
This increase in SRAM leakage current is large enough to contribute a significant fraction of the total SRAM domain power consumption, as shown in Figure \ref{fig_power}.
\textcolor{black}{Despite the increase in leakage current shown in Figure \ref{fig_leakage}, a 2.5 \% decrease in total power consumption was measured in both the core and peripheral domain.}

\subsection{SEU Laser Testing}
The SEU tolerance of the STRV-R1 was verified by introducing SEUs using SPA backside-illumination.
At energy levels greater than 250 pJ, SEUs could be reliable introduced in the standard cells as well as the SRAM macros.
The presence of an SEU can be determined by an output signal provided by the chip, which contains information about discrepancies in each voter of the system.
The triple modular redundancy and majority voter placement scheme introduced in the STRV-R1 design have proven effective.
Occurring SEUs in the core peripherals are immediately detected and corrected within one clock cycle by the integrated feedback, unless the laser pulse aligns with the clock edges in which case an additional clock cycle may be required to correct the SEU.
The triplication of SRAM content and the accompanying scrubbing algorithm ensure reliable correction of any upset data and instructions within an upper time limit of 320 \textmu s.

\section{Conclusion}
In this paper, we showed an SEE tolerant RISC-V implementation with a TMR based protection \textcolor{black}{scheme}. Moreover, a scrubbing algorithm was adapted to the SRAM to prevent the accumulation of SEUs.
The SRAM protection scheme has been identified to have the greatest potential for resource optimization and a reduction in power consumption.
Future research will compare different techniques in the detection and mitigation of radiation induced errors to facilitate the development of a highly SEU tolerant RISC-V based system with reduced resource requirements.
Studies using high-energy X-ray irradiation have demonstrated the suitability of the chosen 65nm CMOS technology to withstand a TID of 500 Mrad. Laser induced fault injections were used to verify the effectiveness of this design and it’s SEU protection techniques.

\end{document}